\documentclass[12pt]{article}
\usepackage{pic04}
\usepackage{hyperref}
\usepackage{url}
\usepackage{graphicx}
\usepackage{amssymb}

\newcommand{\pom} {I\!\!P}
\newcommand{\pomsub} {{\scriptscriptstyle \pom}}
\newcommand{\xpom} {x_{\pomsub}}
\newcommand{\apom} {\alpha_{\pomsub}}
\newcommand{\aprime} {\alpha^\prime_\pomsub}
\newcommand{\gv}{\gamma^\star}
\newcommand{\gp}{\gamma p}
\newcommand{\gvp}{\gamma^\star p}
\newcommand{\units}{\,\mathrm}

\newcommand{\gevtwo}{\units{GeV^2}}
\newcommand{\gevmtwo}{\units{GeV^{-2}}}

\newcommand{\ftwod}{F_2^D}
\newcommand{\ftwodthree}{F_2^{D(3)}}

\newcommand{\ftwopom}{F_2^{\pomsub}}

\begin{document}

\title{\bf DIFFRACTIVE SCATTERING}
\author{
Halina Abramowicz        \\
{\em Raymond and Beverly Sackler Faculty of Exact Sciences}\\{\em
School of Physics and Astronomy } \\{\em Tel Aviv University, Tel Aviv, Israel}}
\maketitle

%
%
%
%
%
%
\vspace{4.5cm}
%

\baselineskip=14.5pt
\begin{abstract}
Recent experimental results on inclusive diffractive scattering and on
exclusive vector meson production are reviewed. The dynamical picture
of hard diffraction emerging in perturbative QCD is highlighted.
\end{abstract}
\newpage

\baselineskip=17pt

\section{Introduction}
In hadron-hadron scattering, interactions are classified by the
characteristics of the final states.  In elastic scattering, both
hadrons emerge unscathed and no other particles are produced. In
diffractive dissociation, the energy transfer between the two
interacting hadrons remains small, but one (single dissociation) or
both (double dissociation) hadrons dissociate into multi-particle
final states, preserving the quantum numbers of the associated initial
hadron.  The remaining configurations correspond to inelastic interactions.

The most difficult conceptual aspect of diffractive scattering is to
provide a unique and concise definition. This will not be attempted
here and diffraction will be understood as an interaction between
projectile and target that generates a large rapidity gap between the
respective final states, which is not exponentially suppressed. 

Diffractive interactions are mediated by the exchange of a colorless
object, with quantum numbers of the vacuum. This definition fits very
well the framework of soft interactions, where diffractive scattering
is mediated by the exchange of the universal Pomeron trajectory
($\pom$), introduced by Gribov~\cite{Gribov:1961}. Ingelman and
Schlein~\cite{Ingelman:1985} proposed to use diffractive scattering in
the presence of a large scale to establish the partonic content of the
Pomeron.

In QCD, the candidate for vacuum exchange with properties similar to
the soft Pomeron is two gluon exchange~\cite{Low:1975,Nussinov:1975}. As a
result of interactions between the two gluons, a ladder structure
develops. In perturbative QCD (pQCD), the properties of this ladder
depend on the energy and scales involved in the interaction, implying
its non-universal character. 

Each of the approaches mentioned above leads to definite predictions,
which can be tested in high energy diffractive interactions in the
presence of a hard scale. This has been pursued in $ep$ scattering at
HERA and in $p\bar{p}$ scattering at the Tevatron. The purpose of this talk
is to summarize the recently achieved progress.

\section{Kinematics of hard diffractive scattering}
The variables used to analyze diffractive scattering will be
introduced for deep inelastic $ep$ scattering (DIS). Since DIS is
perceived as a two-step process, in which the incoming lepton emits a
photon which then interacts with the proton target, the relevant
variables can be readily generalized to $p\bar{p}$ interactions.
\begin{figure}[htb]
\begin{center}
\includegraphics[width=0.25\hsize]{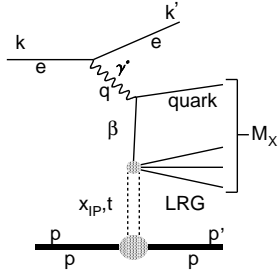}
\caption{\it Schematic diagram for diffractive DIS in $ep$ interactions.} 
\label{fig:dis-diag}
\end{center}
\end{figure}
A diagram for diffractive scattering in DIS, where the diffracted
state is separated from the scattered proton by a large rapidity gap
(LRG), is presented in figure~\ref{fig:dis-diag} and all the relevant
four vectors are defined therein. The usual DIS variables are the
negative of the mass squared of the virtual photon,
$Q^2=-q^2=-(k-k^\prime)^2$, the square of the center of mass energy of
the $\gvp$ system, $W^2=(q+p)^2$, the Bjorken scaling variable,
$x=\frac{Q^2}{2p\cdot q}$, which in the Quark Parton Model constitutes
the fraction of the proton momentum carried by the interacting quark,
and the inelasticity, $y =
\frac{p\cdot q}{p \cdot k}$. In addition to the usual DIS variables,
the variables used to described the diffractive final state are,
\begin{eqnarray}
t &=& (p-p^\prime)^2 \, ,
\label{eq:deft} \\
\xpom &=& \frac{q\cdot (p-p^\prime)}{q \cdot p}
\simeq \frac{M_X^2+Q^2}{W^2+Q^2} \, ,
\label{eq:defxpom} \\
\beta &=& \frac{Q^2}{2q \cdot (p-p^\prime)} = \frac{x}{\xpom}
\simeq \frac{Q^2}{Q^2+M_X^2} \, .
\label{eq:defbeta}
\end{eqnarray}
$\xpom$ is the fractional proton momentum which
participates in the interaction with $\gv$. It is sometimes denoted by
$\xi$. $\beta$ is the equivalent of Bjorken $x$ but
relative to the exchanged state. $M_X$ is the invariant mass of the
hadronic final state recoiling against the leading proton,
$M_X^2=(q+p-p^\prime)^2$.  The approximate relations hold for small
values of the four-momentum transfer squared $t$ and large $W$,
typical of high energy diffraction.

\section{Formalism of diffractive scattering}

To describe diffractive DIS, it is customary to choose the variables
$\xpom$ and $t$ in addition to the usual $x$ and $Q^2$ in the cross
section formula. The diffractive contribution to $F_2$ is denoted by
$F_2^D$ and the corresponding
differential contribution, integrated over $t$, is
\begin{equation}
F_2^{D(3)}=\frac{dF_2^D}{d\xpom} \, , \ \ \ \ \ 
\end{equation}
The three-fold
differential cross section for $ep$ scattering can be written as
\begin{equation}
\frac{d^3\sigma^D_{ep}}{ d\,\xpom d\,x d\,Q^2 }
=\frac{2\pi \alpha^2}{x Q^4} \left[ 1+(1-y)^2\right] 
\sigma_r^{D(3)}(x,Q^2,\xpom) \, ,  \label{eq:f2d4} 
\end{equation}
where
\begin{equation}
\sigma_r^{D(3)} = \ftwodthree -\frac{y^2}{1+(1-y)^2}F_L^{D(3)} \, .
\label{eq:sigr}
\end{equation}
$F_L^{D(3)}$ stands for the diffractive longitudinal structure
function, which may not be small. 
The structure function $F_2$ is related to the absorption cross
section of a virtual photon by the proton, $\sigma_{\gamma^\star p}$.
For diffractive scattering, in the limit of high $W$ (low $x$),
\begin{equation}
F_2^{D(3)}(x,Q^2,\xpom) = \frac{Q^2}{4\pi^2\alpha}
\frac{d^2\sigma^D_{\gamma^\star p}}{ d\,\xpom} \, .
\label{eq:gstarp}
\end{equation}
This relation allows predictions for diffractive scattering in DIS
based on Regge phenomenology applied to $\gvp$ scattering. In fact
many of the questions that are addressed in analyzing diffractive
scattering are inspired by Regge phenomenology as established in soft
hadron-hadron interactions.

\subsection{Regge phenomenology}
The scattering of two hadrons, $a$ and $b$, at squared center of mass
energy $s \gg m^2_{a,b},\, t$, is described by the exchange of the
universal $\pom$ trajectory parameterized as $\apom(t)=\apom(0) + \aprime t$.
The $\pom$ trajectory determines the $s$ dependence of the total
cross section, $\sigma_{tot}\sim s^{\apom(0)-1}$. The
ratio of elastic and diffractive to total cross sections, is expected
to rise like $s^{\apom(0)-1}$. A steep and universal
$\xpom$ dependence of the diffractive cross section is expected,
$d\sigma^D/d\xpom \sim \xpom^{-(2\apom(t)-1)}$. 

Values of $\apom(0)=1.081$~\cite{Donnachie:1992} and $\aprime=0.25
\gevmtwo$~\cite{Donnachie:1984} were derived based on total
hadron-proton interaction cross sections and elastic proton-proton
data.  Recently the $\pom$ intercept has been
reevaluated~\cite{Kang:1998,cudell} leading to a value of
$\apom(0)=1.096 \pm 0.03$. 

The positive value of $\aprime$ implies that the slope of the $t$
distribution is increasing with $\ln s$.  This fact, borne out by the
hadron-hadron and photoproduction data (for a review and new data
see~\cite{zeus-hight}), is known as shrinkage of the $t$ distribution.
It is due to the fact that $\aprime > 0$ and has been explained by
Gribov~\cite{Gribov:1961} as diffusion of particles in the exchange
towards low transverse momenta, $k_T$, with $\aprime \sim 1/k_T^2$
(see also~\cite{Forshaw:1997}).

\subsection{QCD factorization and diffractive partons}
QCD factorization for the diffractive structure function of the
proton, $\ftwod$, is expected to
hold~\cite{Collins:1998,Berera:1996,Trentadue:1994}, while it cannot
be proven for hadron-hadron interactions~\cite{Collins:1998}. $\ftwod$ is
decomposed into diffractive parton distributions, $f^D_i$, in a way
similar to the inclusive $F_2$,
\begin{equation}
\frac{d\ftwod (x,Q^2,\xpom,t)}{d\xpom dt}=\sum_i \int_0^{\xpom} dz 
\frac{d f^D_i(z,\mu,\xpom,t)}{d\xpom dt} \hat{F}_{2,i}(\frac{x}{z},Q^2,\mu) \, ,
\label{eq:fact}
\end{equation}
where $\hat{F}_{2,i}$ is the universal structure function for DIS on
parton $i$, $\mu$ is the factorization scale at which $f^D_i$ are
probed and $z$ is the fraction of momentum of the proton carried by
the diffractive parton $i$. Diffractive partons are to be understood
as those which lead to a diffractive final state. The DGLAP evolution
equation applies in the same way as for the inclusive case. For a fixed
value of $\xpom$, the evolution in $x$ and $Q^2$ is equivalent to the
evolution in $\beta$ and $Q^2$.

If, following Ingelman and Schlein~\cite{Ingelman:1985}, one further
assumes the validity of Regge factorization, $\ftwod$ may be decomposed
into a universal $\pom$ flux and the structure function of the $\pom$,
\begin{equation}
\frac{d\ftwod (x,Q^2,\xpom,t)}{d\xpom dt}= f_{\pom/p}(\xpom,t)
\ftwopom (\beta,Q^2) \, ,
\label{eq:f2pom}
\end{equation}
where the normalization of either of the two components is arbitrary.
It implies that the $\xpom$ and $t$ dependence of the diffractive
cross section is universal, independent of $Q^2$ and $\beta$, and given
by
\begin{equation}
f_{\pom/p}(\xpom,t) \sim \left( \frac{1}{\xpom} \right)^{2\apom(0)-1}
e^{(b_0^{D}-2\aprime \ln \xpom)t} \, ,
\label{eq:pomflux}
\end{equation}
one of the expectations which is subject to experimental tests.

The mechanism for producing LRG is assumed to be present at some scale
and the evolution formalism allows to probe the underlying partonic
structure. The latter depends on the coupling of quarks and gluons to
the Pomeron.

\section{Measurements of $\ftwod$ at HERA}
\label{sec:incl}
\begin{figure}[htbp]
\centerline{\hbox{\hspace{0.2cm}
    \includegraphics[width=6.5cm]{figs/f2d3lpsmxh1.epsi}
    \hspace{0.3cm}
    \includegraphics[width=6.5cm]{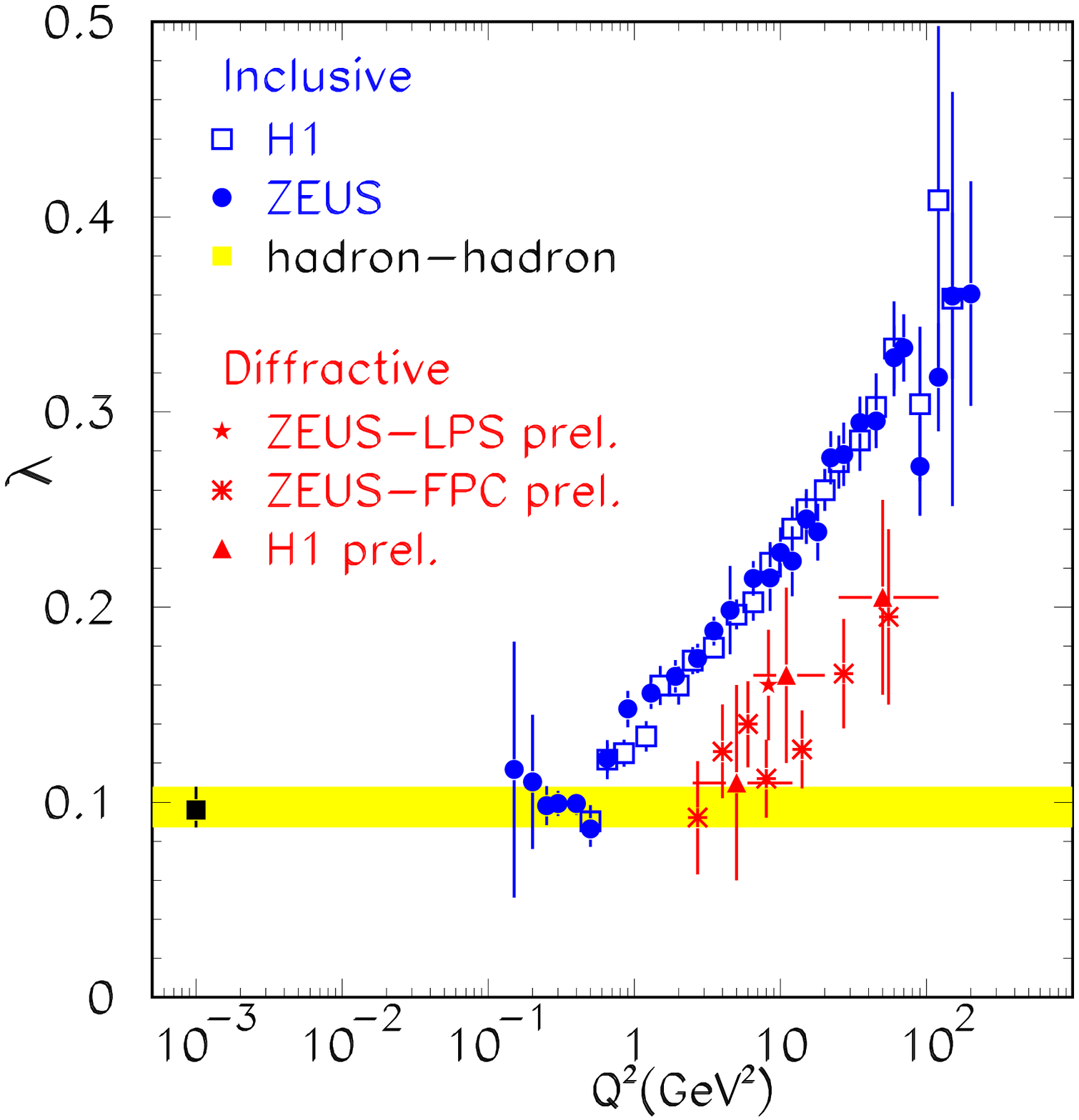} }  }
\caption{\it Left: comparison of $\xpom \ftwodthree$ measured by
H1~\protect\cite{h1f2d} and ZEUS~\protect\cite{zeuslpsf2d,zeusfpc} as
a function of $\xpom$ in overlapping bins of $\beta$ and $Q^2$. Right:
$Q^2$ dependence of $\lambda=\apom(0)-1$ fitted to the measurements of
$F_2$ and $F_2^D$ of the proton, as denoted in the figure.  } 
\label{fig:f2dcomp}
\end{figure}

At HERA the diffractive candidate events are selected either by
requiring a large rapidity gap~\cite{h1f2d,zeusfpc}, or by requiring a
leading proton~\cite{zeuslpsf2d}. The various analyzes differ in the
way the non-diffractive contributions are treated and in the way the
proton dissociative events are subtracted. 
The comparison between the various measurements is shown in
figure~\ref{fig:f2dcomp}.
The $\xpom$ dependence of $\ftwodthree$, in the region of
$\xpom<0.01$, is expected to be dominated by $\pom$ exchange. The
corresponding values of $(\apom(0)-1)$ are shown as a function of
$Q^2$ in figure~\ref{fig:f2dcomp}. A dependence of $\apom$ on $Q^2$
cannot be excluded, however the errors are large enough, so that a
constant value of $\apom(0)$ fits the data. However averaged over the
whole $Q^2$ range the value of $\apom$ is definitely larger than the
intercept of the soft Pomeron. In addition, the diffractive
$\apom(0)-1$ value is only half of that for inclusive $F_2$
measurements, also shown in the figure. This means that the ratio of
diffractive to total $\gvp$ cross sections is constant with $W$.
Those are indications that the connection of diffractive DIS to the
simple soft $\pom$ picture is not straight forward.

\subsection{Diffractive parton distributions}
\begin{figure}[htbp]
\centerline{\hbox{\hspace{0.2cm}
    \includegraphics[width=6.5cm]{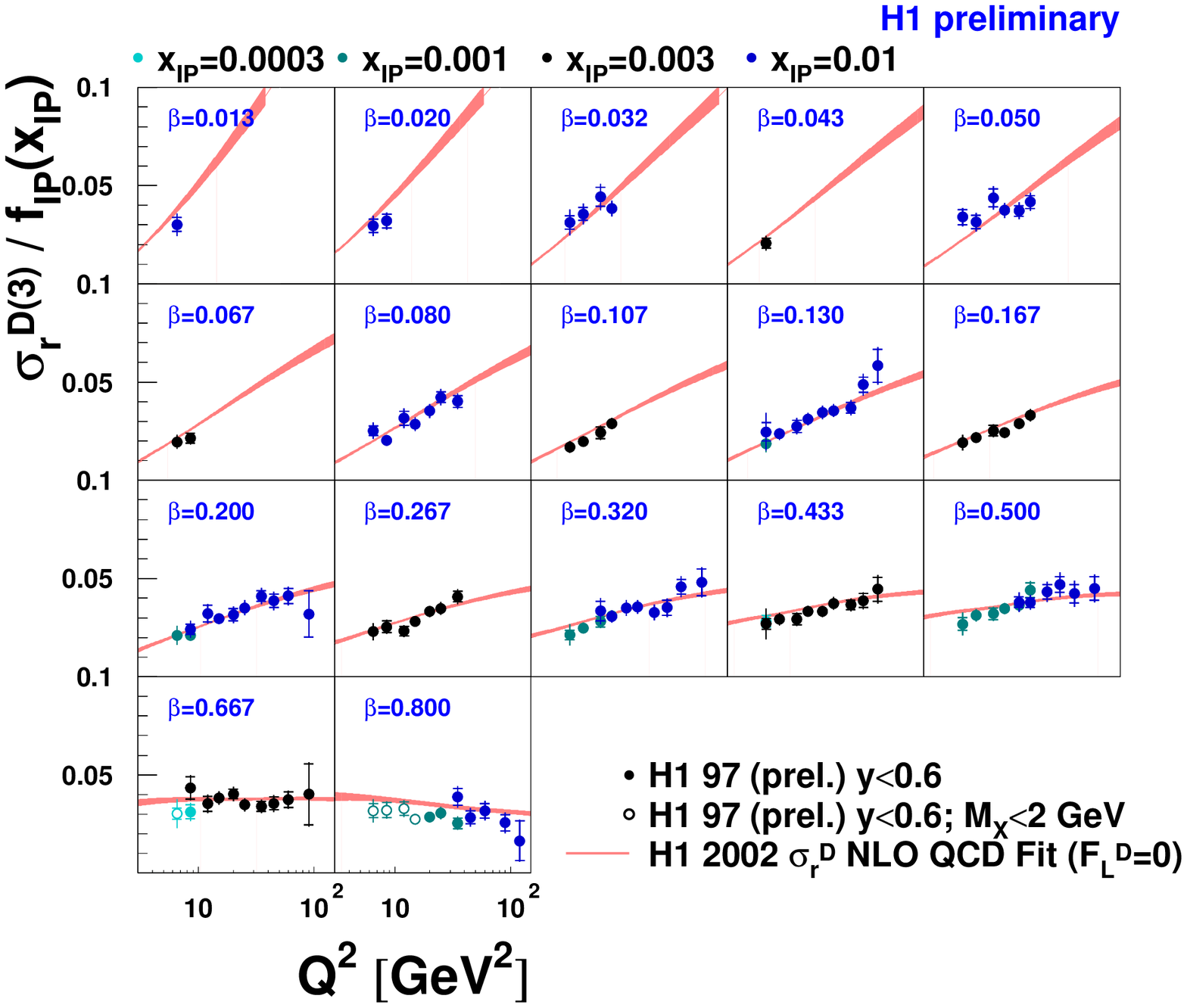}
    \hspace{0.3cm}
    \includegraphics[width=6.5cm]{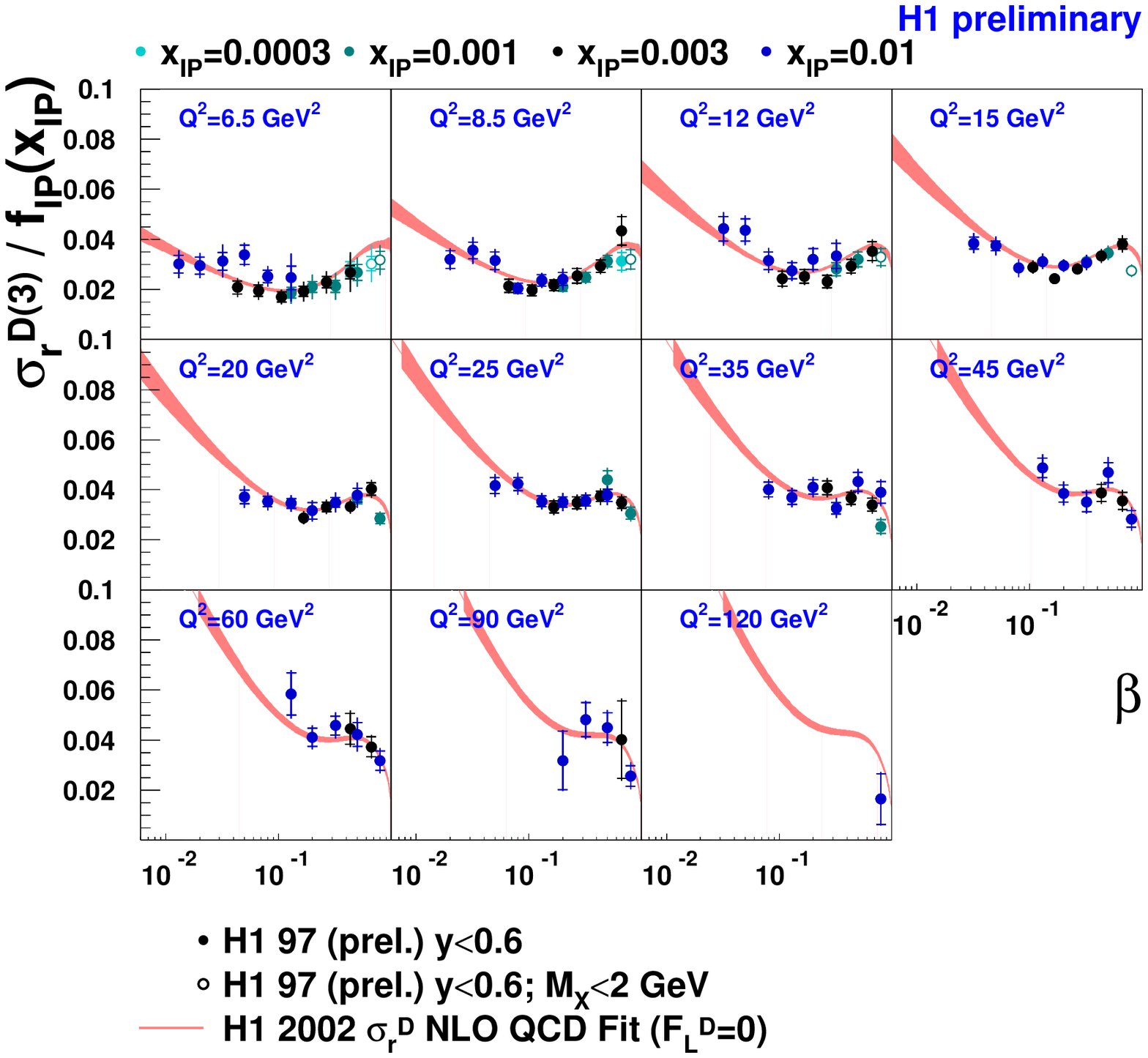} }  }
\caption{\it Comparison of NLO QCD fit with the measurements of
$\sigma_r^D$ after dividing out the flux term. Left, the scaling
violation in bins of $\beta$; right, the $\beta$ distribution in bins
of $Q^2$.  }
    \label{fig:dpdf} 
\end{figure}

The H1 measurements of $\ftwodthree$~\cite{h1f2d}, which cover by far
the largest phase space, have been used to perform a QCD evolution
fit to extract diffractive parton distribution functions (DPDF).
The results of the fit are shown in figure~\ref{fig:dpdf}. 
A very good description of the data is obtained, provided the parton
distributions are dominated by gluons, which carry about 80\% of the
momentum of partons leading to diffractive events. This latest
extraction of DPDFs by H1 is called H1 2002 DPDFs.

\subsection{Tests of QCD factorization}
QCD factorization can be tested in high transverse momenta, $p_T$, jet
production in $\gvp$, $\gp$ and $p\bar{p}$ diffractive production. 

\begin{figure}[htbp]
\centerline{\hbox{\hspace{0.2cm}
    \includegraphics[width=5.5cm]{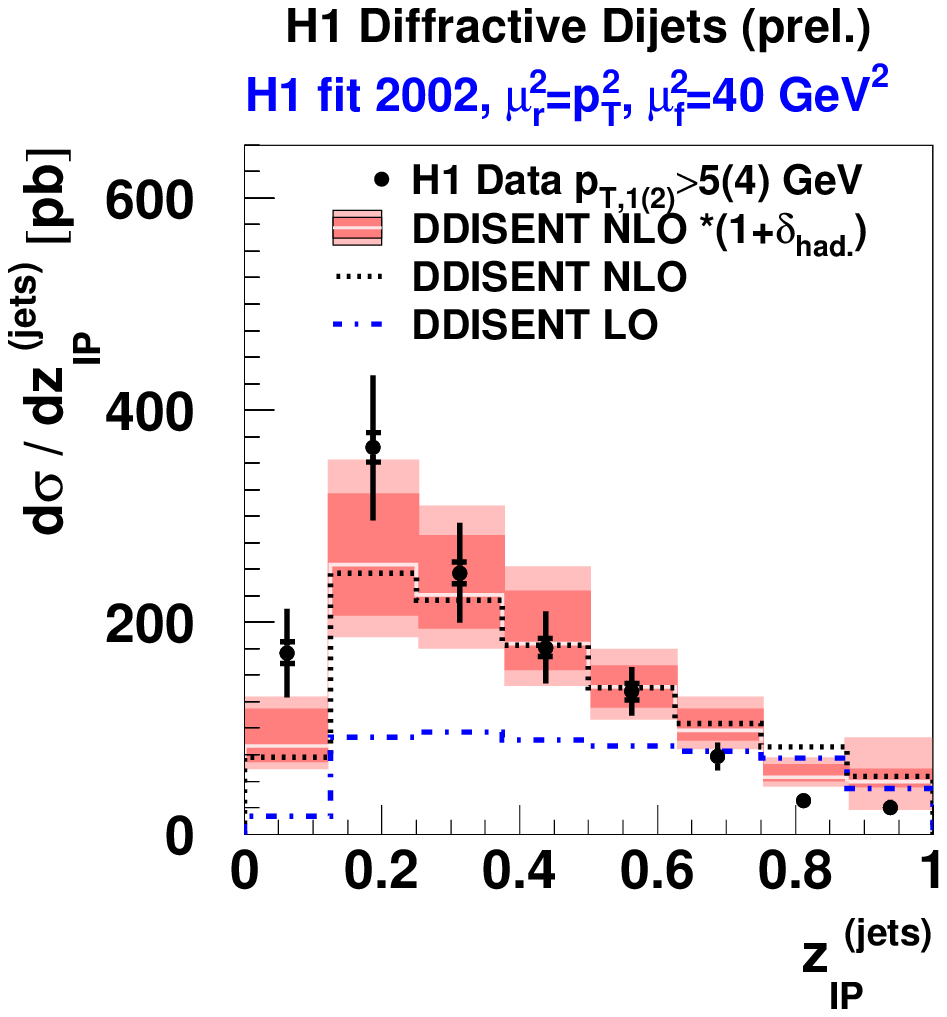}
    \hspace{0.3cm}
    \includegraphics[width=5.5cm]{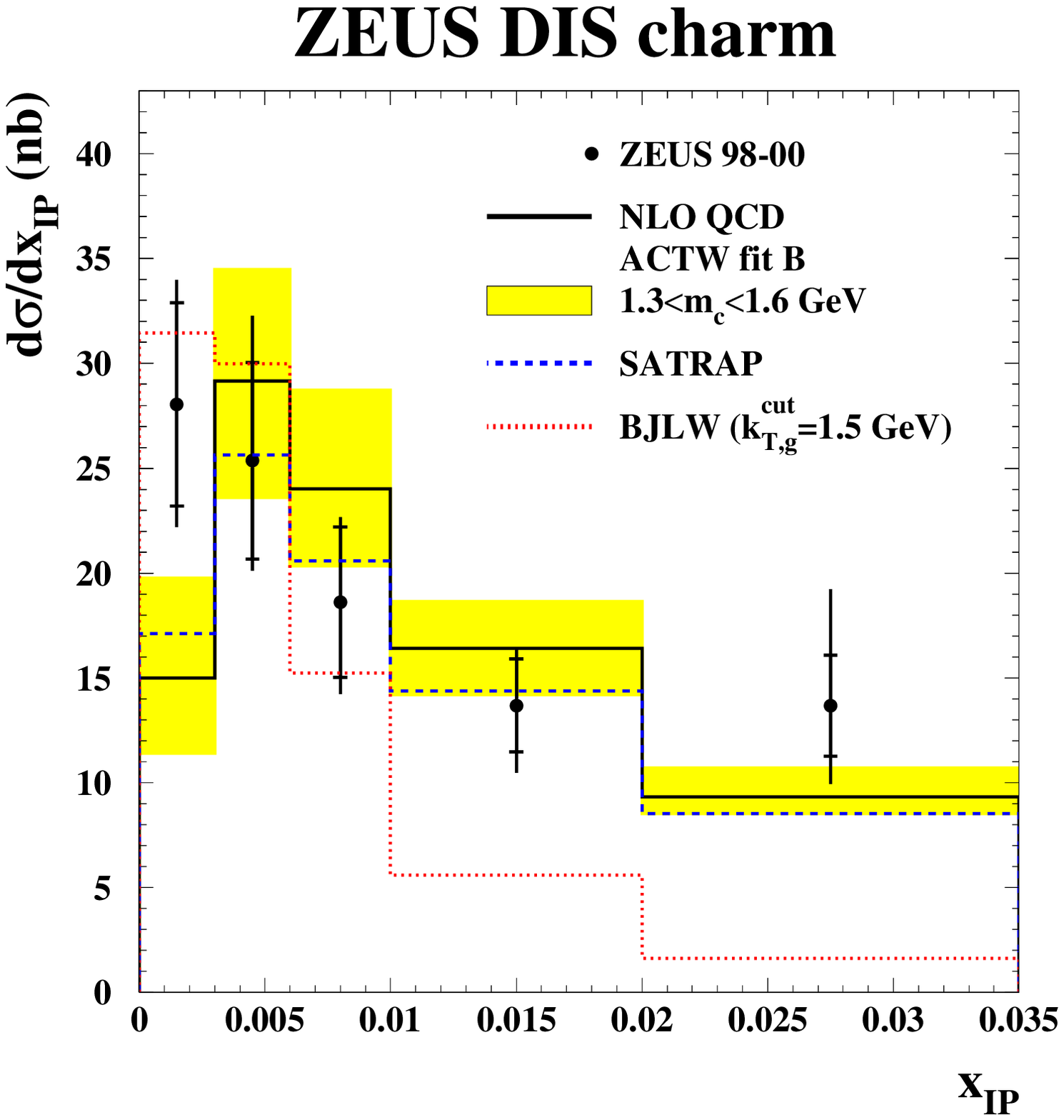} }  }
\caption{\it {Left: distribution of $z_{\pom}^{\mathrm{jets}}$, the
estimate of the diffractive parton momentum fraction of $\pom$, for
dijet production. Right: the $\xpom$ distribution for diffractive
events containing charm in the final state.}}
    \label{fig:jetscc} 
\end{figure}

If factorization holds, the cross section for production of
jets~\cite{h1disjets} and charm~\cite{h1charm,zeuscharm} in $\gvp$ should be
well reproduced by NLO calculations with DPDFs, extracted from
structure function measurements. This is indeed the case as
demonstrated in figure~\ref{fig:jetscc}.

\begin{figure}[htbp]
\centerline{\hbox{\hspace{0.2cm}
\includegraphics[clip,width=5.5cm]{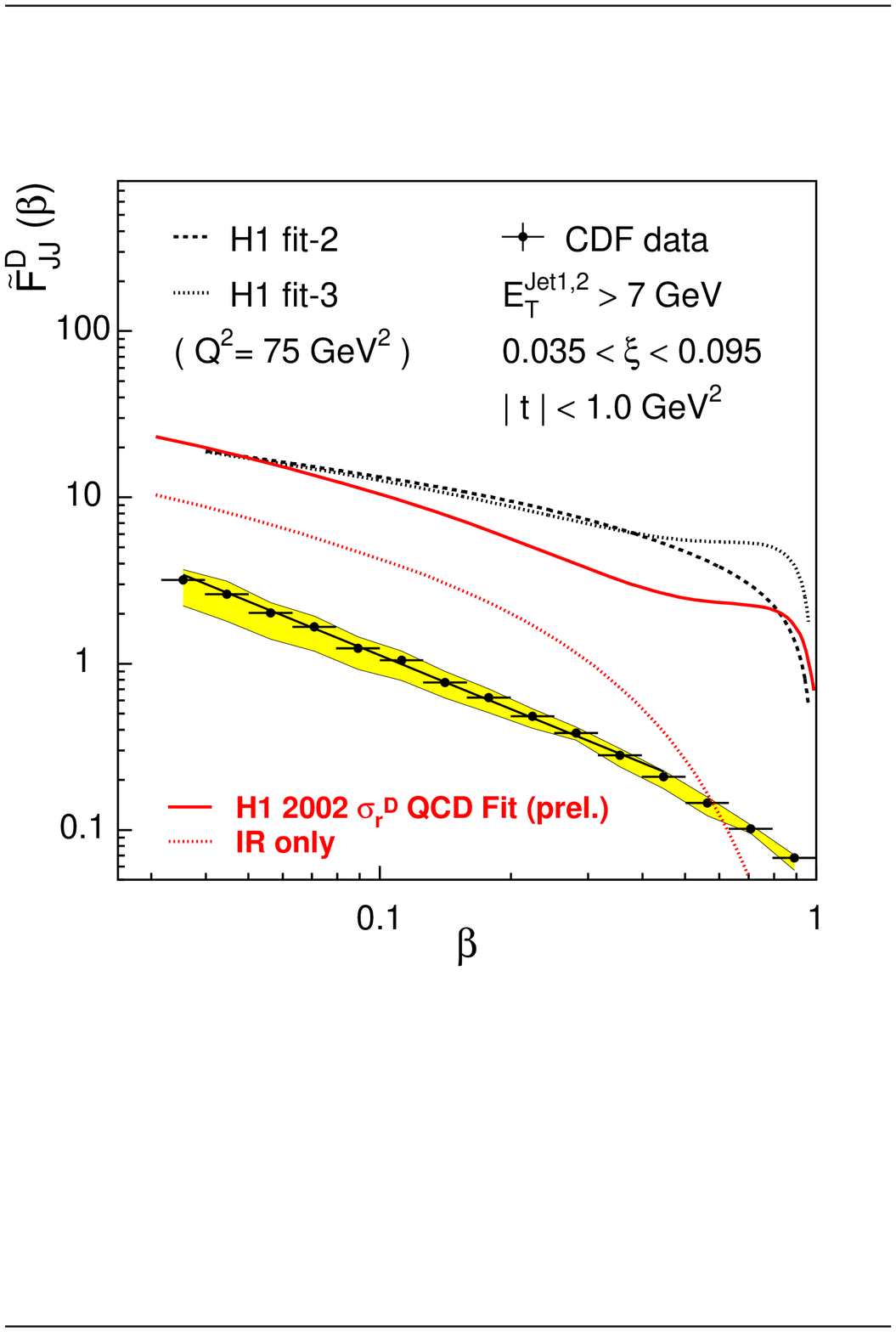}}
    \hspace{0.3cm}
\includegraphics[clip,width=5.5cm]{figs/fjj-dpe.epsi}}
\caption{\it Effective diffractive structure function for dijet
production in $p\bar{p}$ interactions as a function of $\beta$,
compared to expectations of different sets of DPDFs: left, for single
Pomeron exchange, right, double Pomeron exchange.}
\label{fig:fnaljj}
\end{figure}

QCD factorization breaking is observed in hard diffractive scattering
in $p\bar{p}$. Measurements~\cite{fnallps} of two jet
production accompanied by the presence of a leading anti-proton have been
used to extract the effective diffractive structure function for two
jet production $F^D_{JJ}$, which can then be compared with the
expectations from DPDF extracted at HERA. As shown in
figure~\ref{fig:fnaljj}, even for the H1 2002 DPDFs, where the abundance
of gluons is lesser compared to earlier DPDFs~\cite{h1fits:94}, the
expectations are by about a factor 10 above the measurements.
It should be stressed however, that the range of $\xi$ (that is
$\xpom$) covered by the $p\bar{p}$ measurements is beyond the range
probed by the H1 data, from which the DPDFs originate. In extrapolating
the H1 parameterization into this high $\xi$ region, the Reggeon
contribution is estimated to be of the order of 30 to 40\%. 

A lesser QCD factorization breaking is observed for diffractive
dijet production in $p\bar{p}$ events in which both baryons remain
unscathed - the so called double Pomeron exchange (DPE)
process~\cite{fnaldpe}. Compared to expectations, the rate of dijet
production is only by about factor two less abundant. This is shown in
figure~\ref{fig:fnaljj}.

Effects of factorization breaking are also expected for quasi-real
$\gp$ interactions, due to the presence of the resolved photon
component.  The measurements of diffractive dijets by the H1
experiment have been compared to NLO calculations~\cite{klasen} based
on the H1 2002 DPDFs.  
\begin{figure}[htbp]
\centerline{\includegraphics[width=5.5cm]{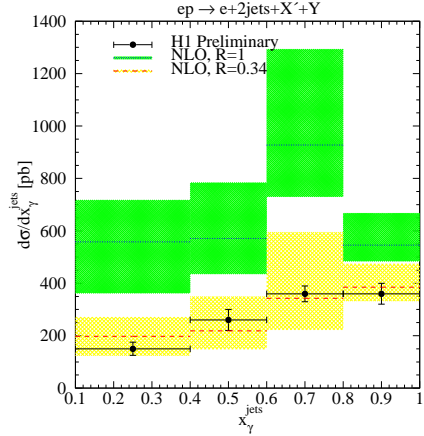}}
\caption{\it Distribution of the fraction of the photon momentum
involved in the hard scattering leading to two jet production,
$x^{jets}_\gamma$, for the diffractive $ep \rightarrow 2
\mathrm{jets}+X'+Y$ reaction, compared to NLO calculations without
($R=1$) or with ($R=0.34$) suppression of the resolved photon
contribution. $X'$ denotes the diffracted system accompanying the two
jets, separated by a large rapidity gap from the proton or its
dissociative state $Y$. }
\label{fig:xgam}
\end{figure}
As shown in figure~\ref{fig:xgam}, a good agreement with data is
obtained if the resolved photon contribution is suppressed
relative to the direct contribution. The factor 0.34 that
multiplies the resolved component in the calculation was motivated by
the recent work of Kaidalov et al.~\cite{kaidalov}.

\section{Unitarity and the dipole picture}
\begin{figure}[htbp]
\centerline{\includegraphics[width=5.cm]{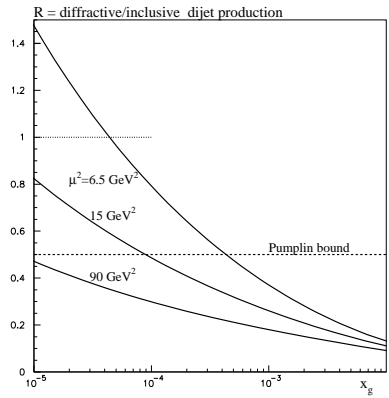}}
\caption{\it The ratio of diffractive to inclusive dijet production
cross section as a function of $x$ of the gluon for different scales
of the hard scattering, for the H1 2002 DPDFs. Also shown is the
unitarity limit, called Pumplin bound.}
\label{fig:xgam2}
\end{figure}

Kaidalov et al.~\cite{kaidalov} investigated what fraction of the gluon
distribution in the proton leads to diffractive final states. The
ratio of diffractive to inclusive dijet production cross sections as a
function of $x$ of the gluon, for different hard scattering scales and
for the H1 2002 DPDFs is presented in figure~\ref{fig:xgam2}.
This ratio should be smaller than 0.5~\cite{pumplin}, while for scales
$\mu^2 =15 \gevtwo$ this limit is exceeded for $x=10^{-4}$. This
indicates that unitarity effects may already be present in diffractive
scattering and may explain why the rise of diffractive scattering
cross section with $W$ is slower than expected (see section~\ref{sec:incl}).

\subsection{The dipole picture}
The dynamics behind diffractive DIS can be easier understood if the
process is viewed in the rest frame of the proton. The virtual photon
develops a partonic fluctuations, whose lifetime is $\tau
=1/2m_px$~\cite{Ioffe:1984}. At the small $x$ typical of HERA, where
$\tau \sim 10 - 100 \units{fm}$, it is the partonic state rather than the
photon that scatters off the proton. If the scattering is elastic, the
final state will have the features of diffraction.

The fluctuations of the $\gv$ are described by the wave functions of
the transversely and longitudinally polarized $\gv$ which are known
from perturbative QCD. Small and large partonic configurations of the
photon fluctuation are present. For large configurations
non-perturbative effects dominate in the interaction and the
treatment of this contribution is subject to modeling.  For a small
configuration of partons (large relative $k_T$) the total
interaction cross section of the created color dipole on a proton target is
given by~\cite{Blaettel:1993,Frankfurt:1999}
\begin{eqnarray}
\sigma_{q\bar{q}p}&=&\frac{\pi^2}{3}r^2\alpha_S(\mu)xg(x,\mu) \, ,
\label{eq:qqp} \\
\sigma_{q\bar{q}gp}&\simeq& \sigma_{ggp}=\frac{9}{4}\sigma_{q\bar{q}p} \, ,
\label{eq:qqgp}
\end{eqnarray}
where $r$ is the transverse size of the color dipole and $\mu \sim
1/r^2$ is the scale at which the gluon distribution $g$ of the proton
is probed. The corresponding elastic cross section is obtained from
the optical theorem. In this picture, the gluon dominance in
diffraction results from the dynamics of perturbative QCD (see
equation~(\ref{eq:qqgp})).

\begin{figure}[htbp]
\centerline{\includegraphics[width=8.5cm]{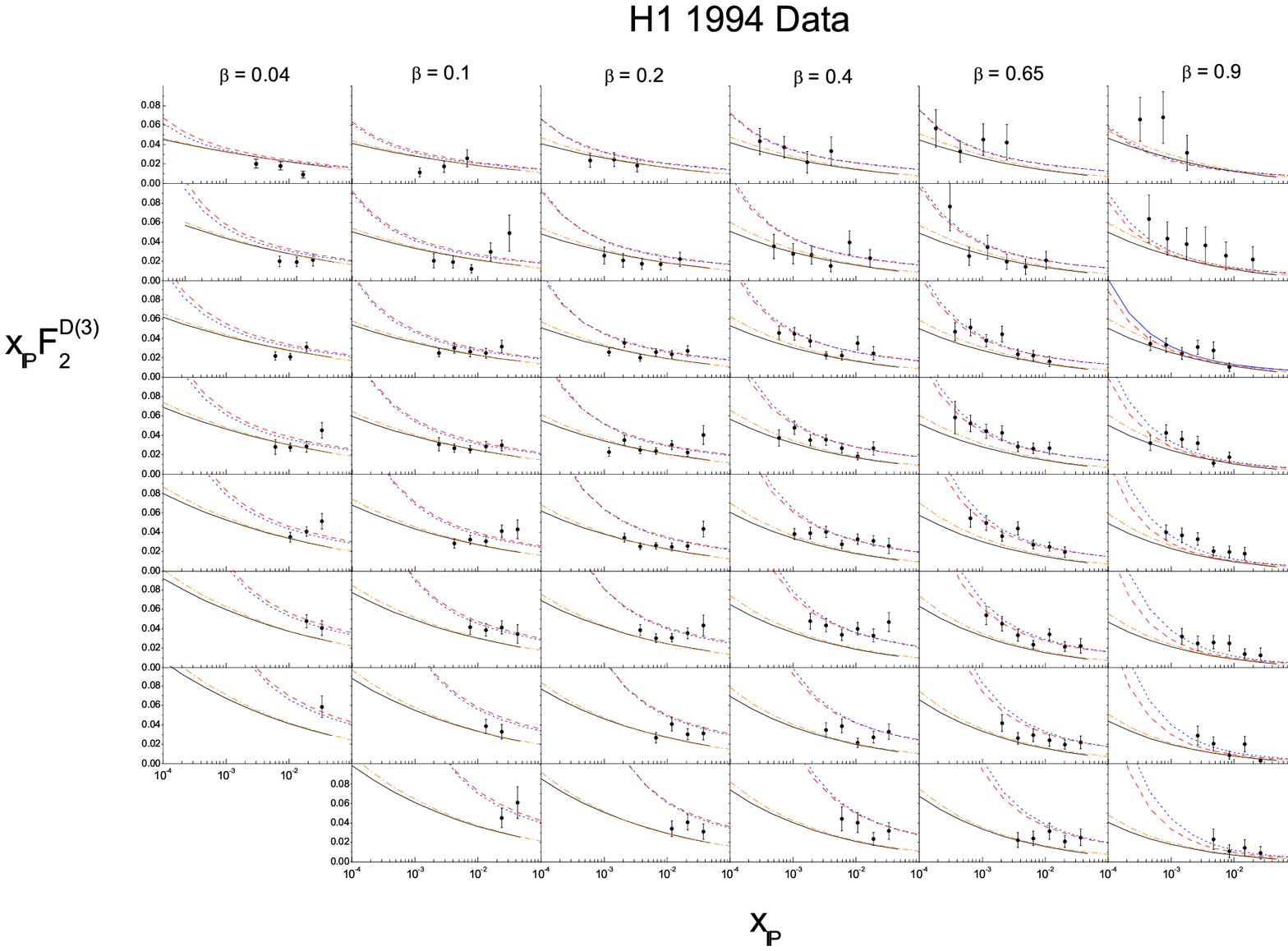}}
\caption{\it Comparison of dipole models 
  from~\protect\cite{forshaw-dipole}, McDermott, Sandapen and
  Shaw~\protect\cite{fks2} (dashed), Forshaw, Kerney and
  Shaw~\protect\cite{fks1} (dotted), Golec-Biernat and
  Wuesthoff~\protect\cite{gbw} (dahed-dotted) and Iancu, Itakura and
  Munier~\protect\cite{iim} (continuous) with $\xpom\ftwodthree$
  measurements.}
\label{fig:dipole}
\end{figure}

Models of diffraction that follow this approach are quite successful
in describing both the inclusive $F_2$ and the diffractive $F_2^D$
measurements, where the former are used to parameterize the
dipole-proton cross section. An example taken from~\cite{forshaw-dipole} is
shown in figure~\ref{fig:dipole}.
It is interesting to note that the two models~\cite{gbw,iim} which
predict a relatively mild increase of $\ftwodthree$ with decreasing
$\xpom$ are the ones which explicitly include effects of unitarization
through saturation of cross sections for large size dipoles. The
saturation scale is increasing with increasing $W$. The dynamical
origin of this form of saturation can be derived from a new form of
QCD matter, called color gluon condensate~\cite{cgc}. In this
approach, the data suggest that the saturation scale $Q_s>1 \gevtwo$
for $x<10^{-4}$.

\section{Exclusive processes in DIS}
The presence of small size $q\bar{q}$ configurations in the photon can
be tested in exclusive vector meson (VM) production as well as for
deeply inelastic Compton scattering. At high energy (low $x$) and in the
presence of a large scale (large $Q^2$ or heavy flavor), these
reactions are expected to be driven by two-gluon exchange.  

A closer look at the theory of exclusive processes in QCD shows that
the two partons taking part in the exchange do not carry the same
fraction of the proton momentum. That makes these processes sensitive
to correlations between partons, which are encoded in the so-called
generalized parton distributions, GPDs~\cite{gpds}. These new
constructs relate in various limits to the parton distributions, form
factors and orbital angular momentum distributions. The motivation
behind studies of exclusive processes is to establish the region of
validity of pQCD expectations and ultimately to pursue a full mapping
of the proton structure, which cannot be achieved in inclusive
measurements.

\subsection{Vector meson production}
\begin{figure}[htbp]
\begin{minipage}{0.47\hsize}
\centerline{\includegraphics[angle=-90,width=0.9\hsize]{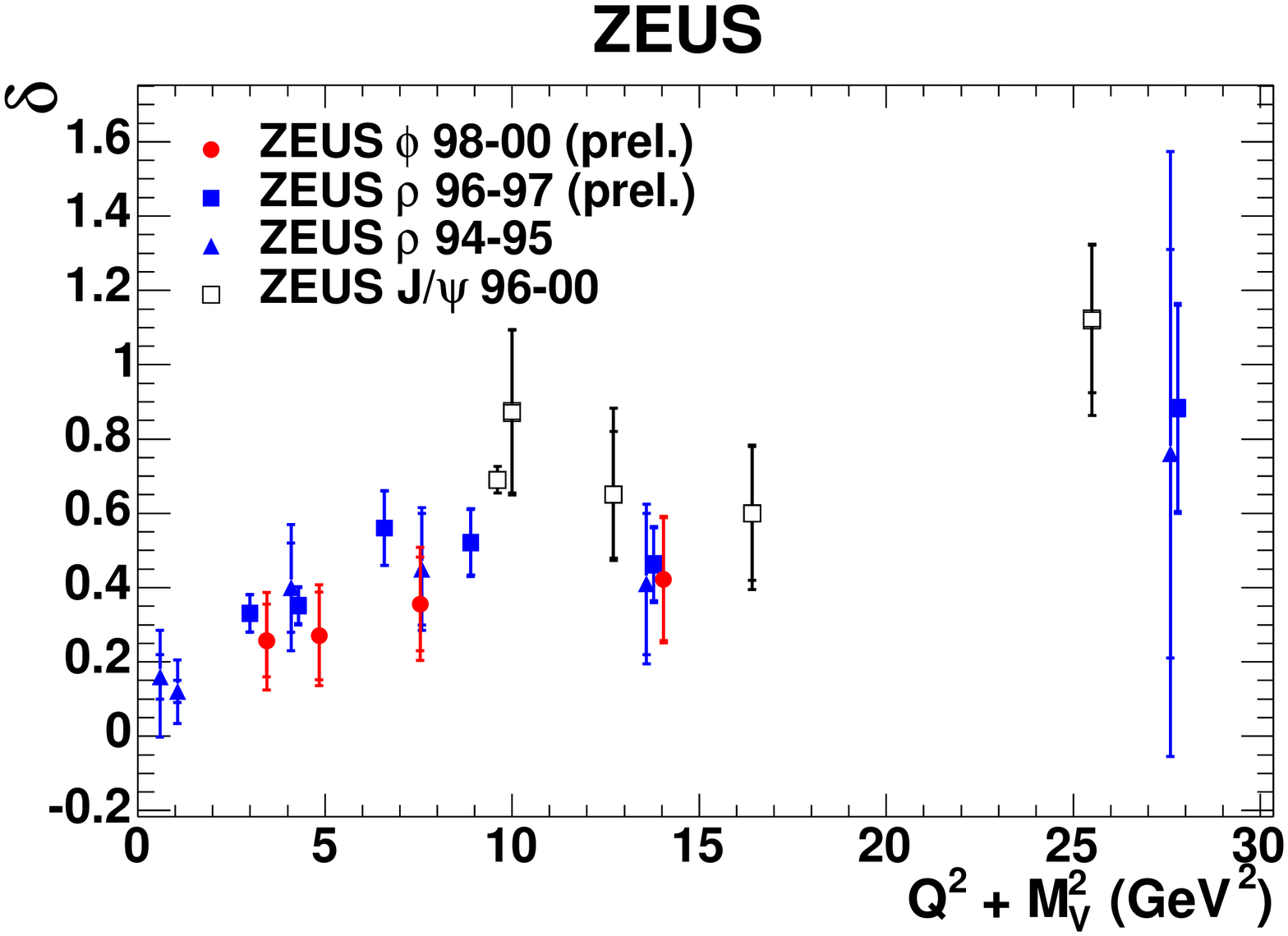}}
\end{minipage}
\begin{minipage}{0.47\hsize}
\centerline{\includegraphics[angle=-90,width=0.9\hsize]{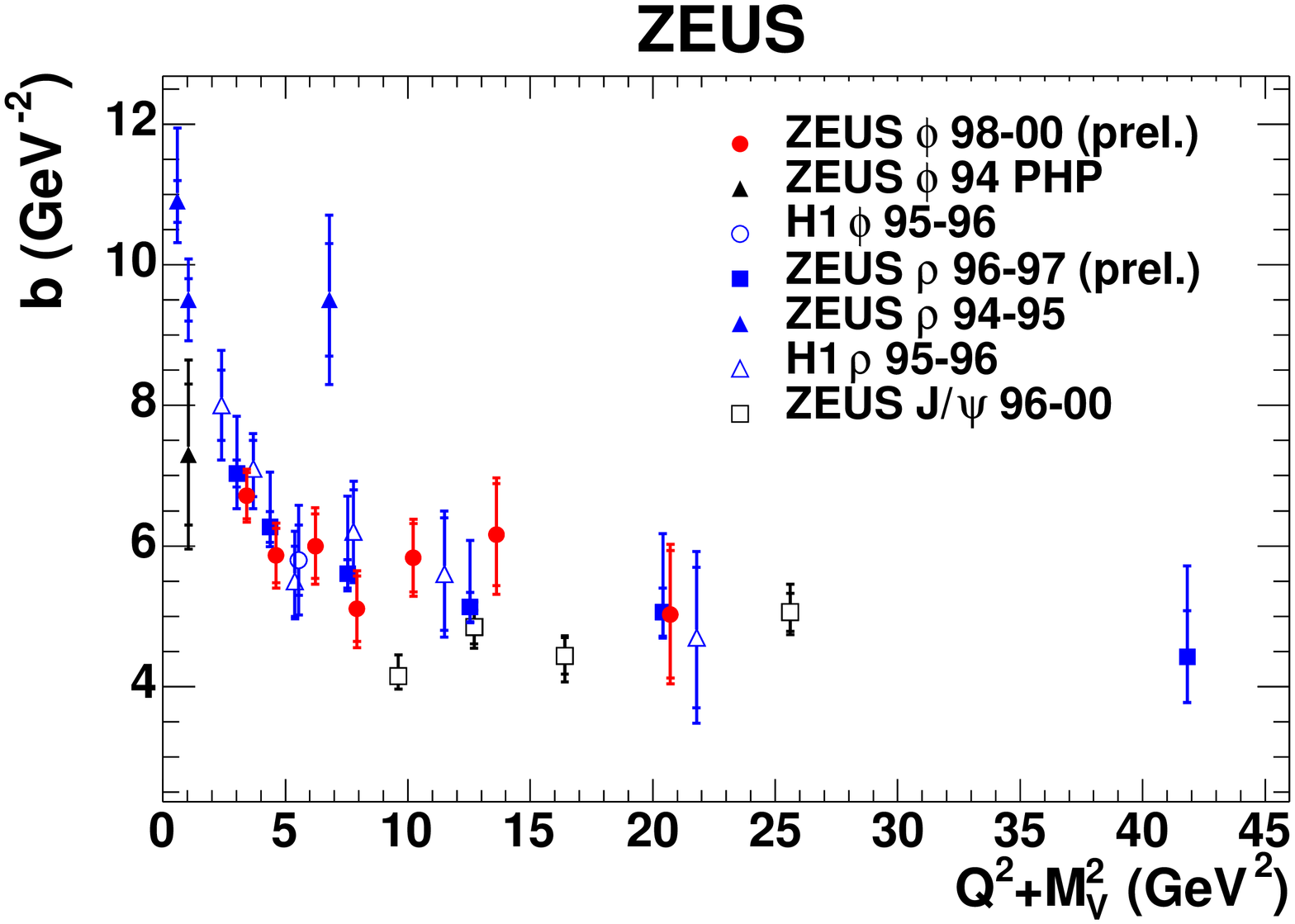}}
\end{minipage}
\caption{\it Left: logarithmic derivatives $\delta=d\log\sigma(\gvp)/d
\log W$ as a function of $Q^2+M_V^2$ for exclusive VM production. 
Right: exponential slope of the $t$ distribution measured for exclusive
VM production as a function of $Q^2+M_V^2$.}
\label{fig:delta}
\end{figure}

The cross section for the exclusive processes is expected to rise with
$W$, with the rate of growth increasing with the value of the hard
scale. A compilation of logarithmic derivatives $\delta=d\log\sigma(\gvp)/d
\log W$, for $\rho$~\cite{zeusrho,h1rho}, $\phi$~\cite{zeusphi,h1phi}
and $J/\psi$~\cite{zeusjpsi,h1jpsi} exclusive production, as a
function of the scale defined as $Q^2+M_{V}^2$, where $M_{V}$ is
the mass of the VM, is presented in figure~\ref{fig:delta}.
With decreasing transverse size of the dipole, the $t$ distribution is
expected to become universal, independent of the scale and of the
VM. The exponential slope of the $t$ distribution, $b$, reflects then
the size of the proton. A
compilation~\cite{zeusrho,h1rho,zeusphi,h1phi,zeusjpsi,h1jpsi} of
measured $b$ values is presented in figure~\ref{fig:delta}.
Around $Q^2+M_V^2$ of about $15 \gevtwo$ indeed the $b$ values become
universal.

\begin{figure}[htbp]
\centerline{\includegraphics[width=5.5cm]{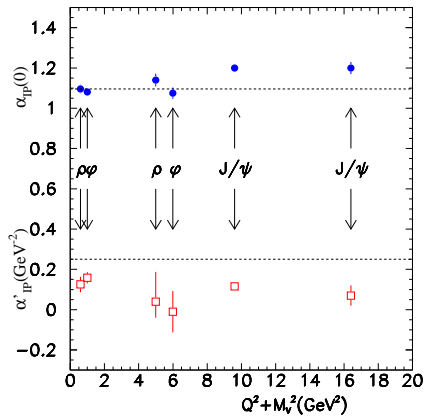}}
\caption{\it Compilation of $\apom(0)$ (dots) and $\aprime$ (open
squares) values, extracted in exclusive VM production, as a function
of $Q^2+M_V^2$.  }
\label{fig:apom}
\end{figure}

Another important manifestation of the perturbative nature of
exclusive processes, related to the universality of the $t$
distribution, is that the slope, $\aprime$, of the corresponding Regge
trajectory should become small. The parameters of the effective Regge
trajectory can be determined in the study of the $W$ dependence of
the differential cross section for exclusive processes at fixed $t$.
The results obtained for exclusive vector meson
production~\cite{zeus-hight,zeusrho,zeusphi,zeusjpsi,zeusjpsi0}
are compiled in figure~\ref{fig:apom}.

\subsection{Deeply virtual Compton scattering}
The deeply virtual Compton scattering (DVCS) process, $\gvp
\rightarrow \gp$, has been advocated as one of the exclusive processes
for which theoretical calculations are free of uncertainties due to
hadronic wave function uncertainties~\cite{strikfurt-dvcs}. In
addition, the interference of the DVCS and QED Bethe-Heitler
amplitudes for prompt $\gamma$ production is proportional to the real
part of the QCD amplitude, which in turn is sensitive to GPDs.

\begin{figure}[htbp]
\centerline{\hbox{\hspace{0.2cm}
\includegraphics[clip,width=7.5cm]{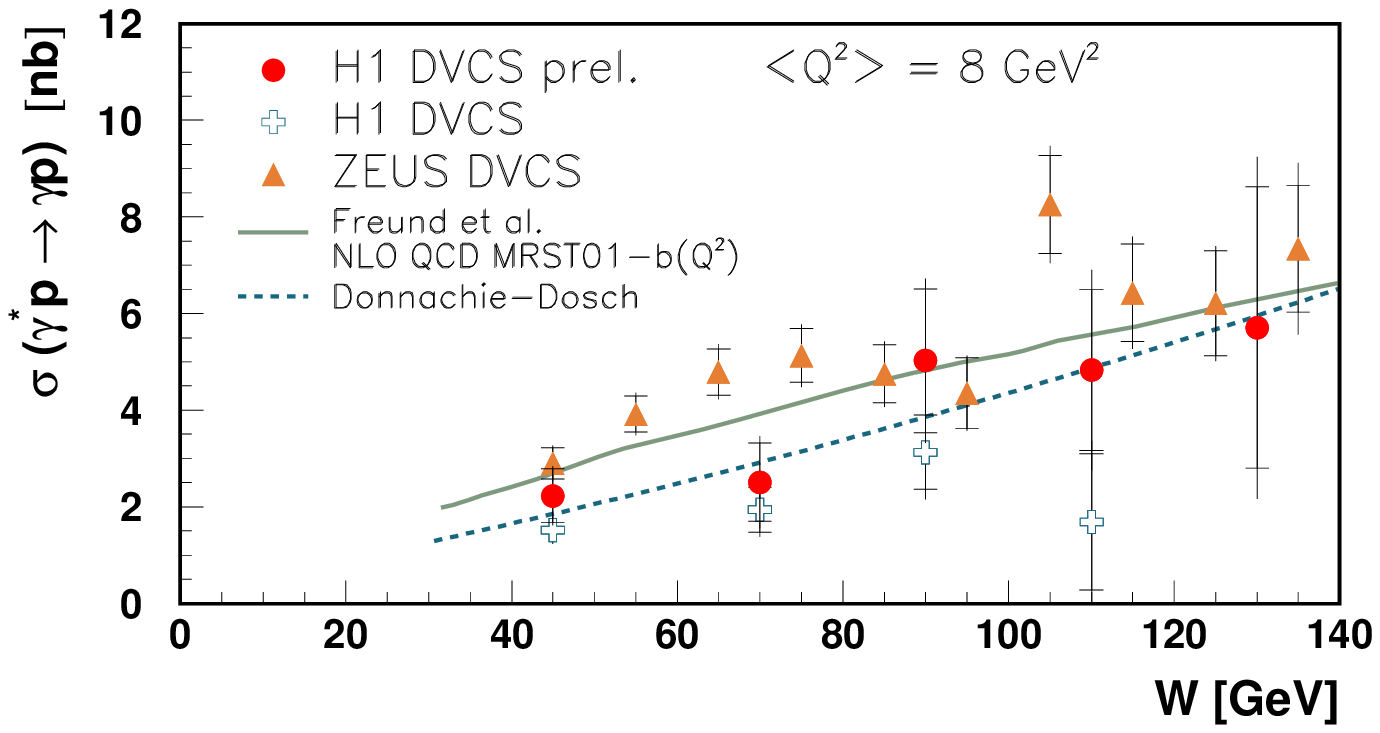}}
    \hspace{0.3cm}
\includegraphics[clip,width=7.5cm]{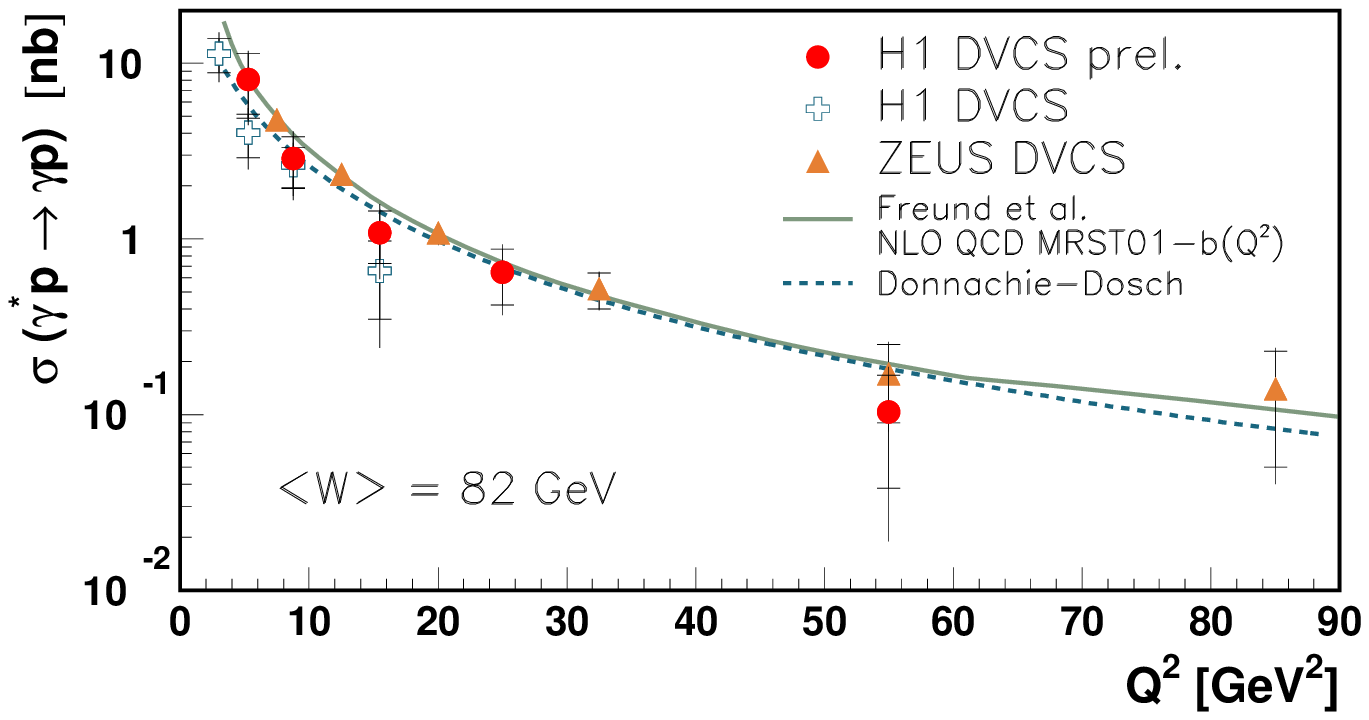}}
\caption{\it The DVCS cross section, $\sigma(\gvp \rightarrow \gp)$ as
a function of $W$ (left) for fixed $Q^2$ and as a function of $Q^2$
for fixed $W$ (right).}
\label{fig:dvcs}
\end{figure}

The extraction of the DVCS cross section in $ep$ scattering has been
performed by the H1~\cite{h1dvcs} and ZEUS~\cite{zeusdvcs}
experiments. A clear rise of the DVCS cross section with $W$ has
been observed~\cite{h1dvcs,zeusdvcs} as shown in
figure~\ref{fig:dvcs}.
A comparison of the DVCS cross section dependence on $W$ and $Q^2$
with two approaches, one based on the dipole
model~\cite{donachie-dosch} and the other on NLO evolution of
GPDs with postulated initial conditions~\cite{freund}, is
shown in figure~\ref{fig:dvcs}.  A good agreement with data is
obtained.

\subsection{Large $t$ exclusive processes}
Diffractive production of VM or prompt $\gamma$ at large values of $t$
accompanied by proton dissociation, form another class of interactions
which are of interest for understanding the high energy regime of
pQCD. The large value of $t$ accompanied by a large rapidity gap,
suggests the applicability of the leading logarithmic BFKL
dynamics~\cite{bfkl}.

\begin{figure}[htbp]
\begin{minipage}{0.5\hsize}
\centerline{
\includegraphics[clip,width=0.8\hsize,angle=-90]{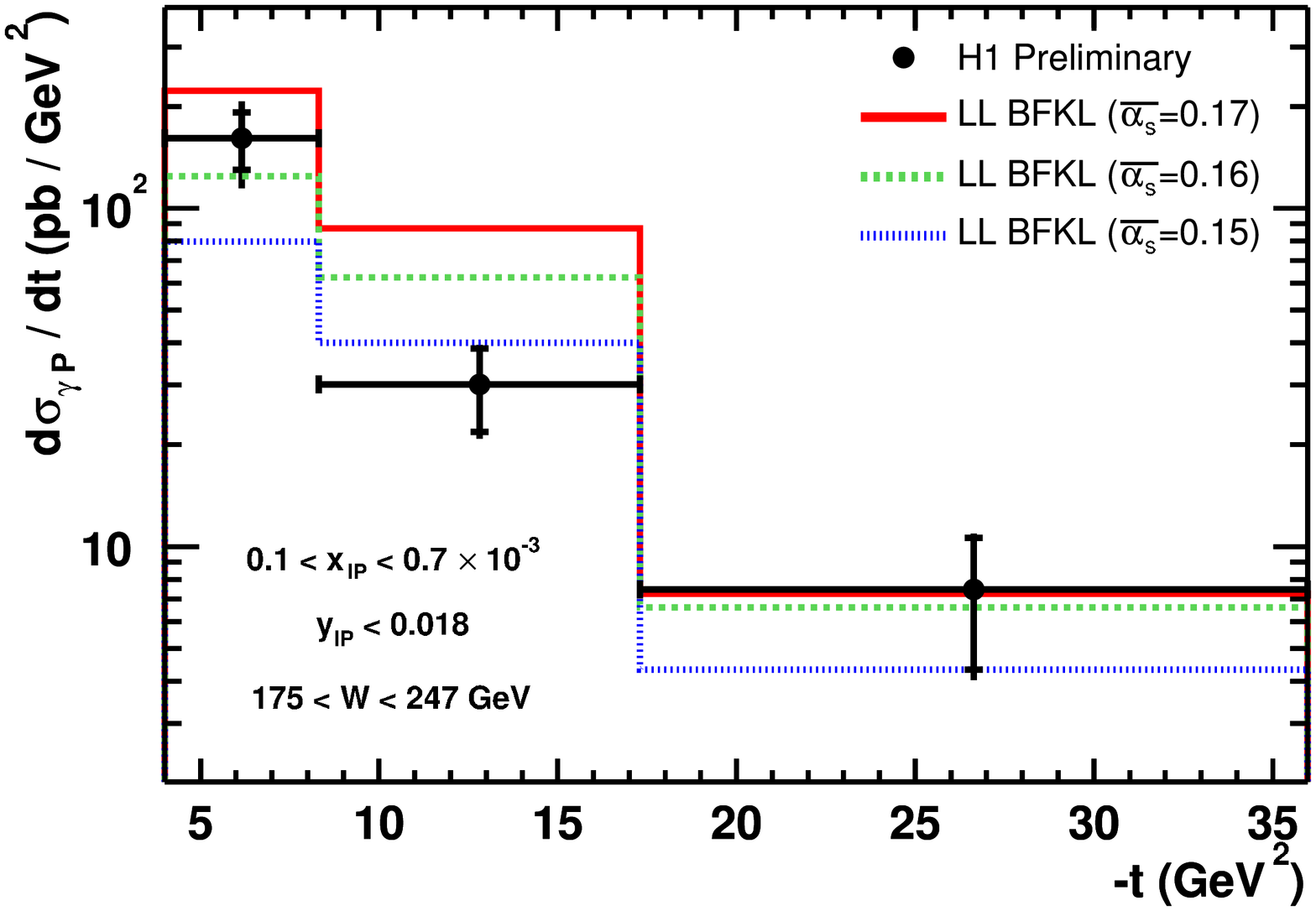}}
\end{minipage}
\begin{minipage}{0.45\hsize}
\centerline{
\includegraphics[clip,width=0.8\hsize]{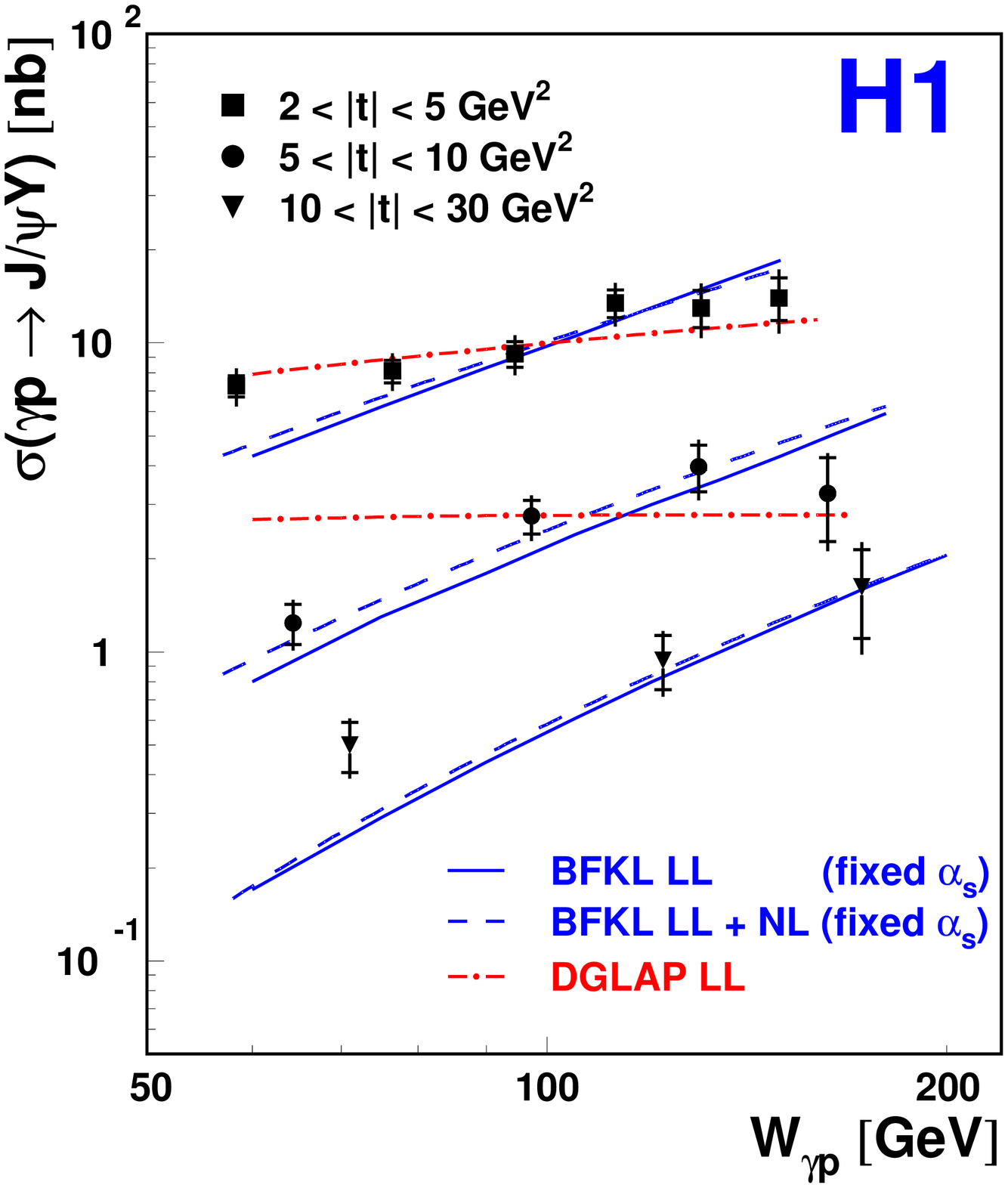}
}
\end{minipage}
\caption{\it Left: distribution of $t$ at the proton vertex in the
process $\gp \rightarrow \gamma Y$. The lines represent theoretical
expectations based on BFKL, as described in the
figure. Right: $W$ dependence of
$\sigma(\gp \rightarrow J/\psi Y)$ in bins of $t$ measured at the
proton vertex. The lines represent theoretical expectations based on
the BFKL or DGLAP dynamics, as described in the figure. }
\label{fig:largetp}
\end{figure}

The $t$ distribution for the process $\gp \rightarrow \gamma
Y$~\cite{h1glt}, where the LRG separates the photon from the
dissociated proton state $Y$ is shown in figure~\ref{fig:largetp}.
The $t$ distribution is well represented by calculations based on
leading logarithmic BFKL approximation~\cite{bfkl}. 

The BFKL approach also describes well the $W$ dependence of the cross
section for exclusive $J/\psi$ production at large $t$, while
expectations based on the leading logarithmic DGLAP dynamics fail to
describe the observed dependence for $t>5 \gevtwo$. This is shown in
figure~\ref{fig:largetp}. This is yet another indication that the
DGLAP dynamics, which is successfully used to describe the
measurements of $F_2$ at HERA~\cite{h1qcd,zeusqcd}, may not be
sufficient to describe all the features of DIS at high energy.

\section{Exclusive states in $p\bar{p}$ interactions}
\begin{figure}[htbp]
\begin{minipage}{0.35\hsize}
\centerline{\includegraphics[clip,width=3.5cm]{figs/chi.epsi}}
\end{minipage}
\begin{minipage}{0.6\hsize}
\includegraphics[clip,width=6.5cm]{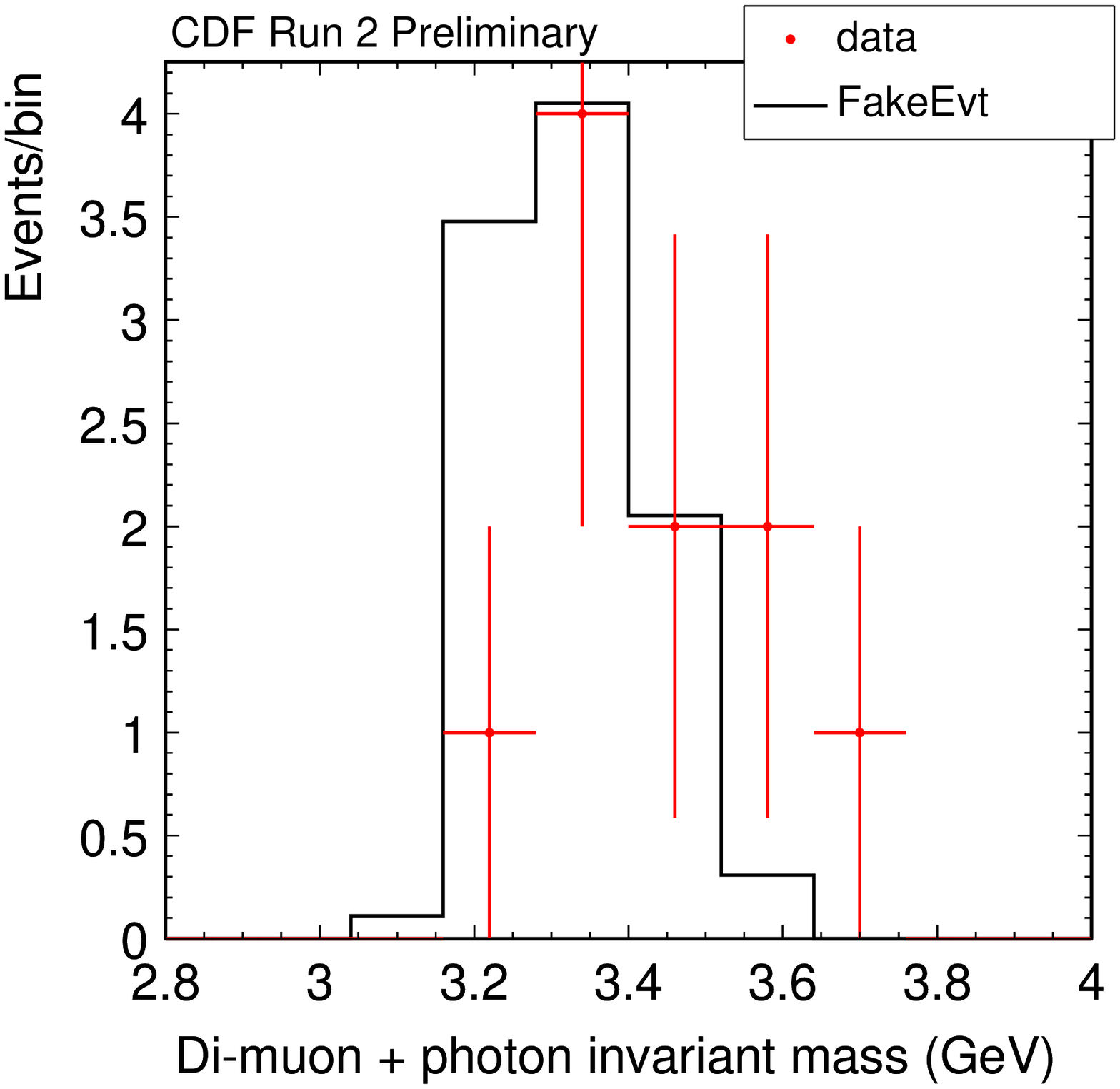}
\end{minipage}
\caption{\it Left: diagram for exclusive $\chi_c^0$ production in
$p\bar{p}$ scattering. Right: invariant mass distribution of the
candidate exclusive $J/\psi(\mu^+\mu^-) \gamma$ system. }
\label{fig:higgs}
\end{figure}

The exclusive diffractive production of the Higgs boson has been
proposed~\cite{higgs-lhc} as a potential background-free method to search
for the light Higgs at LHC. A process similar to exclusive Higgs
production is the exclusive $\chi^0_c$ production, for which at the
Tevatron the cross section is predicted to be about $600
\units{nb}$~\cite{higgs-lhc}. The diagram corresponding to the proposed 
diffractive process is shown in figure~\ref{fig:higgs}.
The CDF experiment~\cite{cdfchi} has searched for the exclusive
process $p\bar{p} \rightarrow p+J/\psi +\gamma +\bar{p}$ and the
invariant mass of the dimuon-photon system for candidate events is
shown in figure~\ref{fig:higgs}. Under the assumption that all found
events originate from exclusive $\chi^0_c$ production, the measured
cross section was estimated to be $50 \pm 18 (\mathrm{stat}) \pm 39
(\mathrm{syst}) \units{pb}$, which is far from the expected number.

\section{Acknowledgements}
I am very thankful to my colleagues from the HERA and FNAL experiments
for their help in collecting the material for this talk. I would also
like to acknowledge the hospitality of the Max Planck Institute in
Munich, where this writeup was prepared, and the Humboldt Foundation
for making my stay in Munich possible. The research was partly
supported by the Israel Science Foundation.

\end{document}